# Climate Change Policies for the XXIst Century: Mechanisms, Predictions and Recommendations

Igor Khmelinskii, FCT, DQF and CIQA, Universidade do Algarve, Campus de Gambelas, Faro, Portugal, ikhmelin@ualg.pt, and

Peter Stallinga, FCT, DEEI and CEOT, Universidade do Algarve, Campus de Gambelas, Faro, Portugal, pjotr@ualg.pt

***Abstract*** — Recent experimental works demonstrated that the Anthropogenic Global Warming (AGW) hypothesis, embodied in a series of Intergovernmental Panel on Climate Change (IPCC) global climate models, is erroneous. These works prove that atmospheric carbon dioxide contributes only very moderately to the observed warming, and that there is no climatic catastrophe in the making, independent on whether or not carbon dioxide emissions will be reduced. In view of these developments, we discuss climate predictions for the XXIst century. Based on the solar activity tendencies, a new Little Ice Age is predicted by the middle of this century, with significantly lower global temperatures. We also show that IPCC climate models can't produce any information regarding future climate, due to essential physical phenomena lacking in those, and that the current budget deficit in many EU countries is mainly caused by the policies promoting renewable energies and other AGW-motivated measures. In absence of any predictable adverse climate consequences of carbon dioxide emissions, and with no predictable shortage of fossil fuels, we argue for recalling of all policies aimed at reducing carbon dioxide emissions and usage of expensive renewable energy sources. The concepts of carbon credits, green energy and green fuels should be abandoned in favor of productive, economically viable and morally acceptable solutions.

***Keywords*** — anthropogenic global warming, IPCC climate models, Little Ice Age, economic crisis.

## I. IPCC CLIMATE SCENARIOS ARE ERRONEOUS

THE terrestrial climate system is quite complex. A complete theory needs to account for the phenomena taking place on the Sun, which provides the energy in the form of electromagnetic radiation that puts the terrestrial climate engine in motion, and the atmosphere and the oceans, that redistribute the energy, with biosphere and humanity affecting the amount of solar energy circulating in the climate system. The energy is eventually re-emitted into outer space in the form of infrared (IR) radiation, maintaining the climate system in a dynamic equilibrium. Surface temperature T is one of the most important climatic variables, which affects the distribution and the very existence of all of the diverse life forms on land, including human civilization, dependent for its existence on availability of potable water and agricultural productivity. The existence of amenable surface temperatures depends on the greenhouse effect, caused by gases present in the atmosphere. The most important greenhouse gas is water vapor ($H_2O$), followed by carbon dioxide ($CO_2$), methane ($CH_4$), ozone and halocarbons. These gases absorb IR radiation that would otherwise immediately escape into space, reemitting part of it back towards the surface. Therefore, they provide some additional thermal insulation for the surface, resulting in the average surface temperature that is ca. 32 K (32 ºC) higher than it would be otherwise. Some 30 years ago, when the growth in atmospheric concentration of $CO_2$, which we shall denote as [$CO_2$], caused by increased human use of fossil fuels, has coincided with the growth in average global temperature (global warming), concerns were risen that the observed warming may be caused by $CO_2$, due to increased greenhouse effect, with possible adverse climatic consequences. These concerns led to the creation of the Intergovernmental Panel on Climate Change (IPCC), an entity under the UN, to evaluate risks of climate change caused by human activity. This entity produces reports, the latest of 2007, describing the current state of climate science as perceived by IPCC experts [1]. The paradigm therefore currently used by a significant number of researchers adopts the IPCC conclusions as correct, proceeding to study emissions of greenhouse gases, which IPCC deems are causing global warming, and then the past and possible future effects of the allegedly continuing global warming upon different aspects of ecosystems and economic activities and their mitigation, see for example [2] – [5].

### A. *Experiment vs the IPCC theory*

As it happens, the only evidence that ever indicated that global warming may be caused by growing atmospheric [$CO_2$] was that of the climate models. These are implemented in software, running on supercomputers, and first make use of real climate information to tune various parameters present in the models, attempting next to produce predictions/ projections on the future climate based on different scenarios of atmospheric [$CO_2$] evolution and other relevant human-generated perturbations, such as changes in land use. Climate models used by the IPCC to asses the effect of atmospheric carbon dioxide on climate predict an average global warming of 3.5 K per doubled [$CO_2$]. Such doubling may happen by the end of this century if carbon dioxide emissions continue unabated at the same rate [1]. The consequences of such a large temperature increase might be quite serious, as the temperature rise would be even more pronounced in the temperate and cold climatic zones, with some undesirable consequences for humanity.

However, any and every climate model belongs to the domain of theories; while a theory only remains valid while there is no experimental evidence that contradicts it. Recently, several publications evaluated the dependence between the surface temperature anomaly and the anomaly in the long-wave (infrared) plus short-wave (ultraviolet and visible) radiation escaping from Earth to the outer space [6], [7]. Here anomalies are defined as excursions from what is considered a "normal" value, usually an average over a certain time period. This dependence may be expressed by a linear relation, with the regression coefficient that we shall denote by α, which indicates the increase in escaping radiation for every degree of temperature rise. The authors of [6], [7] place the experimental value of α, obtained from relevant satellite data, at between 4.5 and 8 $Wm^{-2}K^{-1}$. On the other hand, the IPCC models uniformly produce values that are negative, the average for the set of climate models considered being -2.3 $Wm^{-2}K^{-1}$ [6]. We note an obvious contradiction between the AGW hypothesis and the experiment, as regards the regression coefficient α, both in its numeric value and, even more importantly, its sign (positive *vs* negative).

We shall shortly demonstrate that the coefficient α is of critical importance in evaluating the effect of any perturbation on the terrestrial climate, be it caused by growing atmospheric [$CO_2$], changes in the solar activity, presence of volcanic ash in the upper atmosphere, or any other factor. The reason is that without feedbacks it is fairly easy to evaluate the effect $\Delta T_0$ on the average global temperature T of any perturbation Ω. This effect is obtained assuming that we isolated this perturbation Ω from any other phenomena taking place in the climate system. On the other hand, the practical outcome ΔT of a perturbation is determined by the product of $\Delta T_0$ with the feedback parameter β, $\Delta T = \Delta T_0 \times \beta$, the latter uniquely dependent on the regression coefficient α. This climate feedback parameter β describes the entire complexity of the multiple interconnected phenomena defining the response of the climate system to a perturbation, and is therefore notoriously difficult to obtain theoretically, as evidenced by more than 20 years of the IPCC numerical modeling research.

Continuing, the values of the climate feedback coefficient, β, that exceed unity, as predicted by the IPCC climate models, correspond to amplification of perturbations by the climate system, with the value of β = 2.7 (corresponding to the model value of α = -2.3 $Wm^{-2}K^{-1}$) yielding a predicted temperature change for doubled [$CO_2$] of ΔT = 2.4 K, instead of $\Delta T_0$ = 0.9 K for the no-feedback case [6]. The various model values of β, as obtained in each of the individual climate models, range from about 1.7 to 5.6 or more, and universally exceed unity, predicting an inherently unstable climate system that significantly amplifies any and every perturbation. Contrary to model predictions, the experimental values of β = 0.3...0.6 correspond to a reduction of perturbations by an inherently stable climate system that significantly reduces any perturbation, resisting the imposed changes. These experimental β = 0.3...0.6 yield ΔT values for doubled [$CO_2$] between 0.3 and 0.5 K [6], [7]. These ΔT values, estimated from measurements performed on the real terrestrial climate system, are an order of magnitude below the predictions of the IPCC climate models, which range between 1.5 and 5 to 7 degrees of the average temperature increase for doubled [$CO_2$], with the most probable value of 3.5 K, as reported by IPCC in 2007 [1]. We therefore conclude that the Anthropogenic Global Warming (AGW) hypothesis, embodied in the IPCC climate models, has nothing in common with the real climate, as it fails to predict the value of the crucial parameter of the climate system, namely the climate feedback coefficient, β, which universally defines the response of the climate system to perturbations. Given this verdict, we must expressly disregard every and all of the other results and predictions/projections/scenarios of these models, and every and all of the recommendations based on the aforementioned results and projections/scenarios. Indeed, these models can't be trusted even for qualitative trends, as they miss the correct sense (model amplification, β > 1, *vs* experimental reduction of perturbations, β < 1) of the climate feedback coefficient. We also conclude that the terrestrial climate system is inherently stable, having the climate feedback coefficient β well below unity, contrary to IPCC predictions, and that, for this very reason, there is no imminent climatic catastrophe in the making, anthropogenic (man-made) or otherwise.

Having eliminated [$CO_2$] as the primary and defining driver of current climatic evolution, for complete lack of evidence, we will look at other factors affecting climate, in order to both interpret the warming of the last century and provide a climate forecast for the future. Thus, carbon dioxide can only contribute to the climate change in this century with a very moderate warming, significantly lower than the 0.74 K recorded during the last century. Indeed, of these 0.74 K, knowing the [$CO_2$] increment during the XXth century, and the value of the climate feedback parameter, we can only attribute between 0.1 and 0.2 K to the anthropogenic climate change caused by growth in the atmospheric carbon dioxide, product of fossil fuel combustion. Of these two estimates, the lower value of 0.1 K should be closer to the reality, corresponding to the more realistic higher value of α, as explained below. The remaining warming has therefore been caused by other factors, which may include changes in solar activity, land use, and concentrations of atmospheric aerosols.

### *B. Solar Climate Theory vs Traditional Meteorology and IPCC models*

There is a more fundamental reason allowing to conclude that IPCC climate models can give no insight into the future of the terrestrial climate. The relevant evidence is provided by the applied research by P. Corbyn and co-workers [8], who are consistently producing accurate long-range forecasts of weather and extreme weather events, mostly for UK and Ireland, based on their Solar Climate (SC) theory [8]. Their qualitative approach correlates external factors (EFs), such as solar activity, solar wind and solar magnetic field, and the effects of weak lunar magnetic fields, with past extreme weather events, and then looks for the same set of EFs to reproduce in the near future, and thus predicts future weather up to 12 months in advance, including the extreme weather events, indicating precise date, location and type of event, with 85-90% success rates. Such accurate predictions can't systematically occur by chance, as demonstrated by the inability of traditional models used by meteorologists to predict weather for more than 10 days in advance – confirmed by internet search for weather forecasts – and by their complete inability to produce long-range forecasts, traditionally attributed to the stochastic nature of the terrestrial climate system.

The long-range predictive success of the SC theory, *vs* the failure of traditional meteorology, thus tells us the following:

(1) the terrestrial climate system is deterministic, and not stochastic, its current state uniquely determined by the acting EFs; indeed, as even such rare conditions as extreme weather events are uniquely determined by the state of the EFs, then the entire climate system is tightly controlled by these EFs;

(2) physical phenomena essential for weather and climate prediction are missing from both traditional meteorological weather models and IPCC climate models, notably the interactions of the terrestrial magnetosphere and upper atmosphere with the solar wind, solar short-wave radiation, solar and lunar magnetic fields, with photochemistry and chemistry of ozone and other species present in the upper atmosphere not to be forgotten;

(3) this missing physics is essential for understanding weather and climate, as traditional weather models drift away from the real terrestrial climate system already on the time scale of 2 weeks, which would not happen if we were using complete models and comprehensive input data to model and predict the global weather/climate system;

(4) uniqueness of the current state of the climate system, tightly defined by the EFs, eliminates the possibility of internal instabilities and transitions to radically different states of the global climate system (climate catastrophes), phenomena invoked by IPCC to advocate immediate [$CO_2$] reduction measures;

(5) the fact that the extreme weather events are faithfully reproduced after more than half a century, given that one of the main periodicities naturally arising in the SC theory has the duration of 60 years [8], shows that the EFs are more important for weather and climate than any anthropogenic changes in the climate system that have occurred within the time interval between the two repetitions of a given event, including the rapid growth in [$CO_2$] that occurred during the second half of the XXth century [1], which have not affected the predictability of recent extreme weather events [8]. Note that the 60-year period is well apparent in the record of global sea surface temperatures (see Fig. 1);

(6) periodicity of the extreme weather events and other weather and climate patterns inferred from the SC theory shows that their frequency can't be growing with global warming, contrary to what IPCC is stating in its scenarios and predictions [1], and to what is constantly repeated by media on the daily basis.

Now, not knowing what the EFs are and how and to what extent they affect weather and climate, both traditional meteorologists and IPCC modelers are unable to build models that faithfully reproduce behavior of the real terrestrial climate system. The consequence of this inability for the short-term predictions of the meteorological models is that their predictive ability is lost as soon as the EFs change significantly, which does happen within a couple of weeks. On the other hand, the predictive ability of the IPCC climate models is absent in any time interval they consider, as the EFs unknown to and unaccounted for by the modelers, along with the incorrectly parameterized known physics (as otherwise the models fail to produce the [$CO_2$] warming effects expected by the modelers) will make any prediction of such incomplete and incorrect models drift away from the state of the real climate system, to an arbitrary extent, already on the time scales much shorter than those required for climate predictions. The incorrect parameterization mentioned follows, for example, from the fact that the models fail to reproduce both the sign and the value of the regression coefficient $\alpha$, resulting ion the [$CO_2$] global warming effect deduced from the experimental data being by a factor of 10 smaller than that obtained from the IPCC climate models [1], [6]-[7], as we already discussed in the point *A*.

## II. Climate in this Century

### *A. Phenomenological approach*

Looking at the global temperature record of the last 1.5 centuries, we readily note that warming has not been monotonous. As an example, consider the sea surface temperatures, shown in Fig. 1 and plotted using the Hadley Centre data [9]. Indeed, we note in the monthly global average sea surface temperature record that apart from a general warming trend (with the linear regression coefficient corresponding to a warming of only 0.48 K per century during the time period considered – this shows that continents warm faster than the oceans, the latter lagging behind) we observe a periodic contribution, with a period of about 60 years. The three maxima of this periodic contribution had occurred approximately in 1880, 1940 and 2000, whereas the two minima in 1910 and 1970. In fact, newspapers in the seventies were full of predictions of an imminent new ice age and global freezing, issued by leading climate experts, including some of the present-day advocates of the AGW hypothesis, inspired by a distinct cooling tendency that has lasted for about 30 years, perfectly discernible in the sea, land and global surface temperatures. Note that similar periodic oscillations appear in other temperature records, global as well as regional.

Comparing the two warming periods between 1910-1940 and between 1970-2000, we note that the linear temperature trends are exactly the same. This constitutes yet another proof, if that were needed, that the AGW hypothesis is wrong. Indeed, the fossil fuel consumption in the second half of the XXth century exceeded that of the first half by the factor of 5, with the anthropogenic contributions to atmospheric carbon dioxide equally increasing by at least the factor of 5 between the first and the second half of the XXth century [10]. Were the warming really caused by growth in atmospheric [$CO_2$], as stipulated by the AGW hypothesis, the period of 1970-2000 would have produced at least 5 times as much warming than the 1910-1940 period, contrary to the experimental data shown in Fig. 1, where we see exactly the same amount and rate of warming in these two time periods.

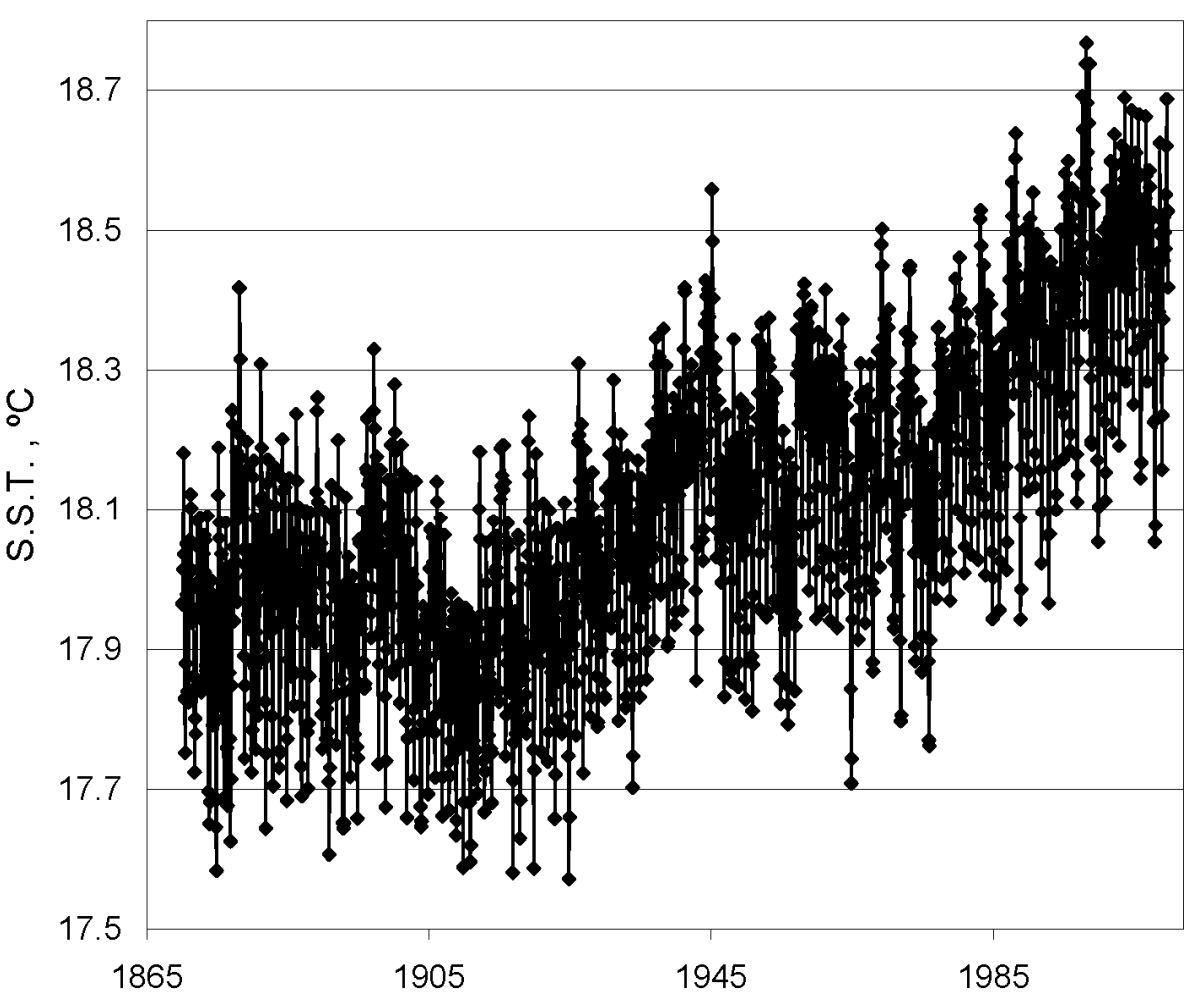

Fig. 1. Global monthly average sea surface temperature from 1870 to 2009 [9].

Therefore, at present the climate system is in the cooling phase of the 60-year oscillation, which will predictably last till about 2030. The current cooling trend has caught the AGW climate modelers by surprise, as they are predicting continuous and ever accelerating warming, as long as atmospheric [$CO_2$] is growing [1]. Obviously, no such thing is happening, demonstrating yet again that the climate change is determined by factors other than atmospheric carbon dioxide. Following this phenomenological treatment, we predict global cooling until about 2030, after which the climate system will be warming once again. Considering the last 150 years of the climate history, for which instrumental temperature record exists (see Fig. 1), we note no tendency of acceleration in the warming rates, therefore, using our phenomenological model, we predict that the Earth can warm during the XXIst century by the same amount as it has during the XXth century, that is, by about 0.7 K, in stark contrast to the catastrophic scenarios produced by IPCC/AGW modelers based on their incomplete, distorted and erroneous models.

### B. *Solar Activity approach*

As we already noted, all of the energy that sets the terrestrial climate machine moving comes from the Sun. This makes the Sun and the phenomena such as periodic changes in the elements of the Earth's orbit around the Sun the most important contributors to the climate change. The solar activity has been monitored by astronomers for hundreds of years, in the simple form of counting sunspots. Fig. 2 shows the yearly sunspot count record for the last 300 years.

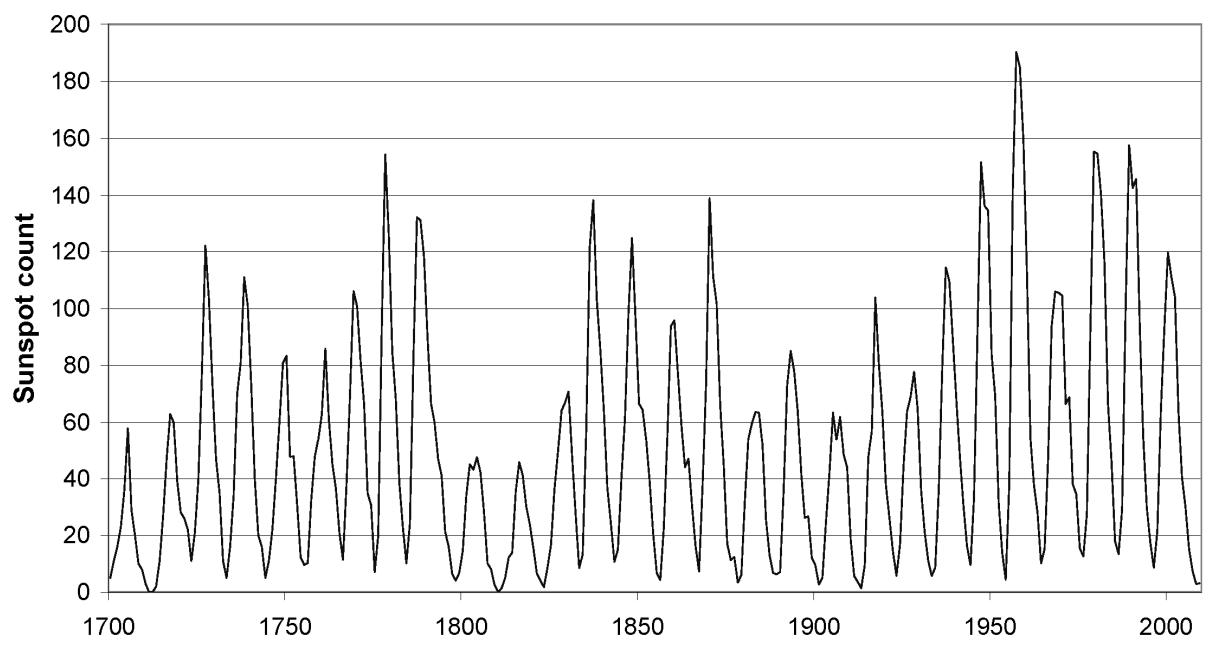

Fig. 2. Sunspot count from 1700 to 2009; the periodic variations correspond to the approximately 11-year Shwabe cycle [11].

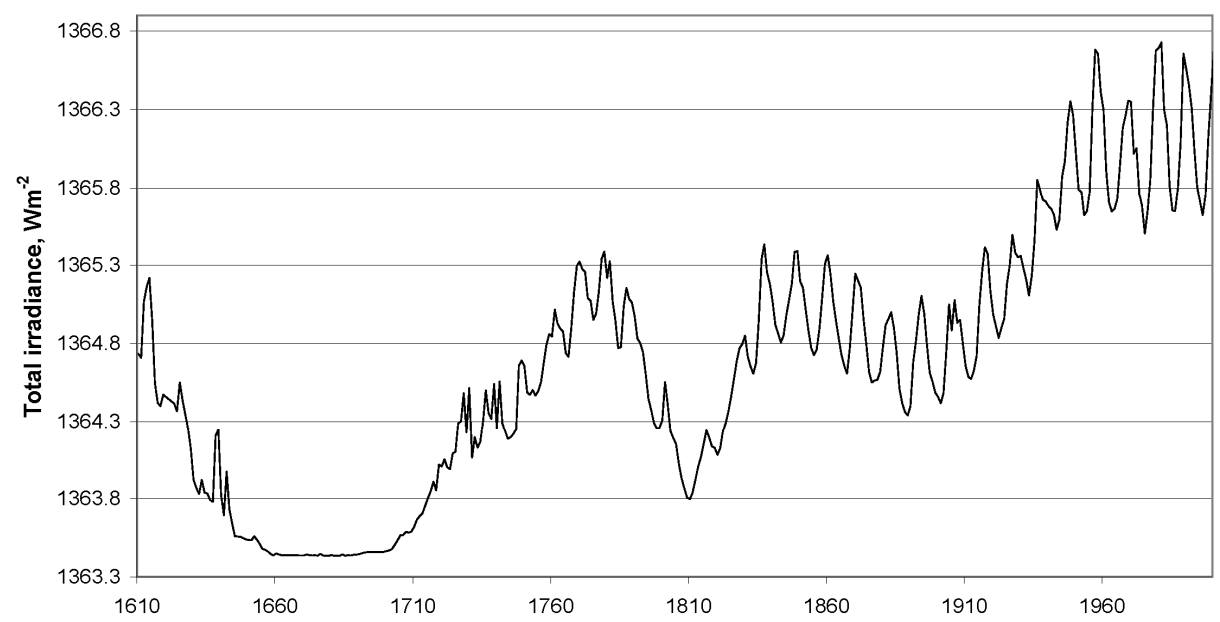

Fig. 3. Total solar irradiance reconstruction since 1610; the minimum centered at ca. 1680 is the Maunder minimum, the minimum centered at ca. 1810 is the Dalton minimum [13].

However, the total solar irradiance (TSI) variations corresponding to the 11-year sunspot cycle shown in Fig. 2 (Schwabe cycle) correspond to only about 0.1% of the TSI value, according to precise satellite measurements performed since 1980. This translates into ca. 1 $Wm^{-2}$, too little to account for the observed climate change [12]. Total solar irradiance was reconstructed for the entire period starting from 1610, taking into account astronomic data on solar-like stars. Such stars were found to emit measurably less in the state similar to the Maunder minimum (see Fig. 3), when the cycling switches off. This allowed to evaluate the slowly changing secular component of the total solar irradiance that could not be yet extracted from the TSI data directly [12]. The resulting reconstruction of the total solar irradiance, tracking the amplitude of the Schwabe cycle, is shown in Fig 3.

Addition of the slowly changing secular component provides a better although still insufficient match between total solar irradiance and climate change, as now the total irradiance change between the Maunder minimum and the contemporary maximum is about 0.24%, which translates into ca. 3 $Wm^{-2}$ and explains at least 30% of the climate change that occurred since 1970, with a larger percentage explainable during preceding periods [12].

A much better correlation between total solar irradiance and climate change was found when irradiance reconstruction tracked the period length of the Schwabe cycle instead of its amplitude [14], [15]. The relevance of the cycle period length as the main parameter determining total solar irradiance is supported by the fact that the Schwabe cycles close to Maunder and Dalton minimums were significantly longer, thus, longer cycles correspond to lower TSI and colder climate. Some of the historic TSI reconstructions attribute almost the totality of the climate change of the XXth century to changes in solar irradiance, conditioned by the remaining uncertainty in the amplitude of secular changes of the solar irradiance, with estimates that vary from 2 to 7 $Wm^{-2}$ [14]–[16].

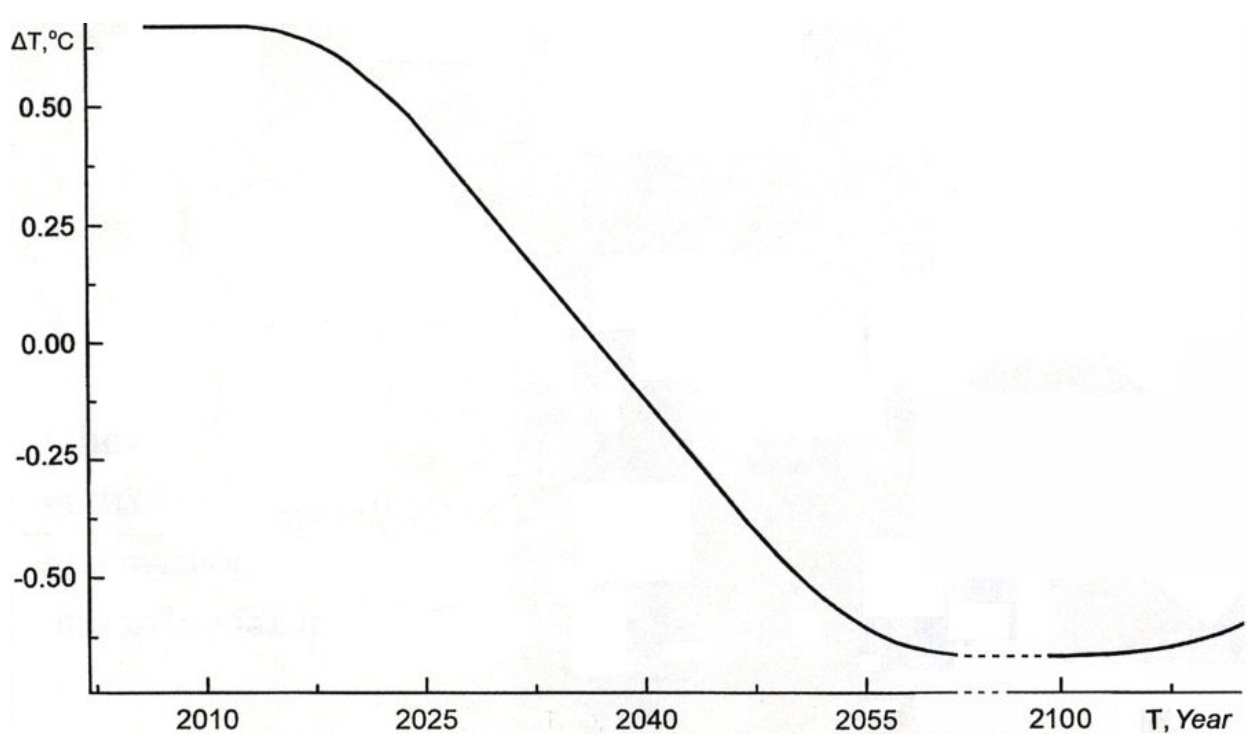

Fig. 4. Climate scenario for the current century: the approaching Little Ice Age, taken from [21], with permission.

Still, there is at least one additional mechanism whereby the solar activity affects our climate system, which complements the TSI changes [17]–[19]. According to this mechanism, galactic cosmic rays (GCR) ionize the lower atmosphere, with the negative ions promoting nucleation of nanoparticles, enabling formation of water drops in the atmosphere at lower partial pressures of water vapor, and therefore increasing the cloud cover. The intensity of GCR, as demonstrated by historic data on $^{14}C$ isotope abundance, increases in low solar activity periods, and decreases in high solar activity periods, when the GCR are swept away by solar wind and deflected by solar magnetic fields and thus reach the Earth at lower intensities. As shown by direct measurements in 1978-1996, the low-altitude cloud cover directly follows the GCR variations. In its turn, GCR variations are anticorrelated to the solar activity as expressed by the sunspot number. Thus, in periods of higher solar activity the GCR intensity and the cloud cover decrease, while the TSI increases, with the two mechanisms acting in synchrony and causing warming [19].

In a separate development, satellite data yielded a trend in the total average solar irradiance at the terrestrial surface from 1983 to 2001 of 0.16 $Wm^{-2}yr^{-1}$, with the total increment of 2.9 $Wm^{-2}$ during the same period [20]. This increment was produced by reductions in the cloud cover (due to reduction of GCR in the period of high solar activity) and reductions in atmospheric aerosols (due to reductions in atmospheric pollution generated by developed countries), and is much larger than any changes in incoming TSI at the outer atmospheric boundary due to solar activity changes during the same period. Recalling the experimental range of the regression coefficient $\alpha = 4.5...8$ $Wm^{-2}K^{-1}$, we thus deduce that the 2.9 W $m^{-2}$ increase in surface irradiance produced $\Delta T = 0.36...0.64$ K of global warming during the same period. This result shows that the more correct value of $\alpha$ should be close to the higher value of 8 $Wm^{-2}K^{-1}$, as otherwise the estimated warming is significantly larger than the about 0.4 K in fact recorded during the last 20 years of the XXth century. Thus, we conclude once more that the climate change observed during the XXth century may be fully explained by the increase in the solar irradiance reaching the terrestrial surface, whereas the role of carbon dioxide and other greenhouse gases has been almost negligible. Given the estimate of $\alpha = 8$ $Wm^{-2}K^{-1}$, we estimate the warming contribution of carbon dioxide of 0.2 K for the current century, provided we don't restrict its emissions, and expecting atmospheric [$CO_2$] doubled by the end of this century.

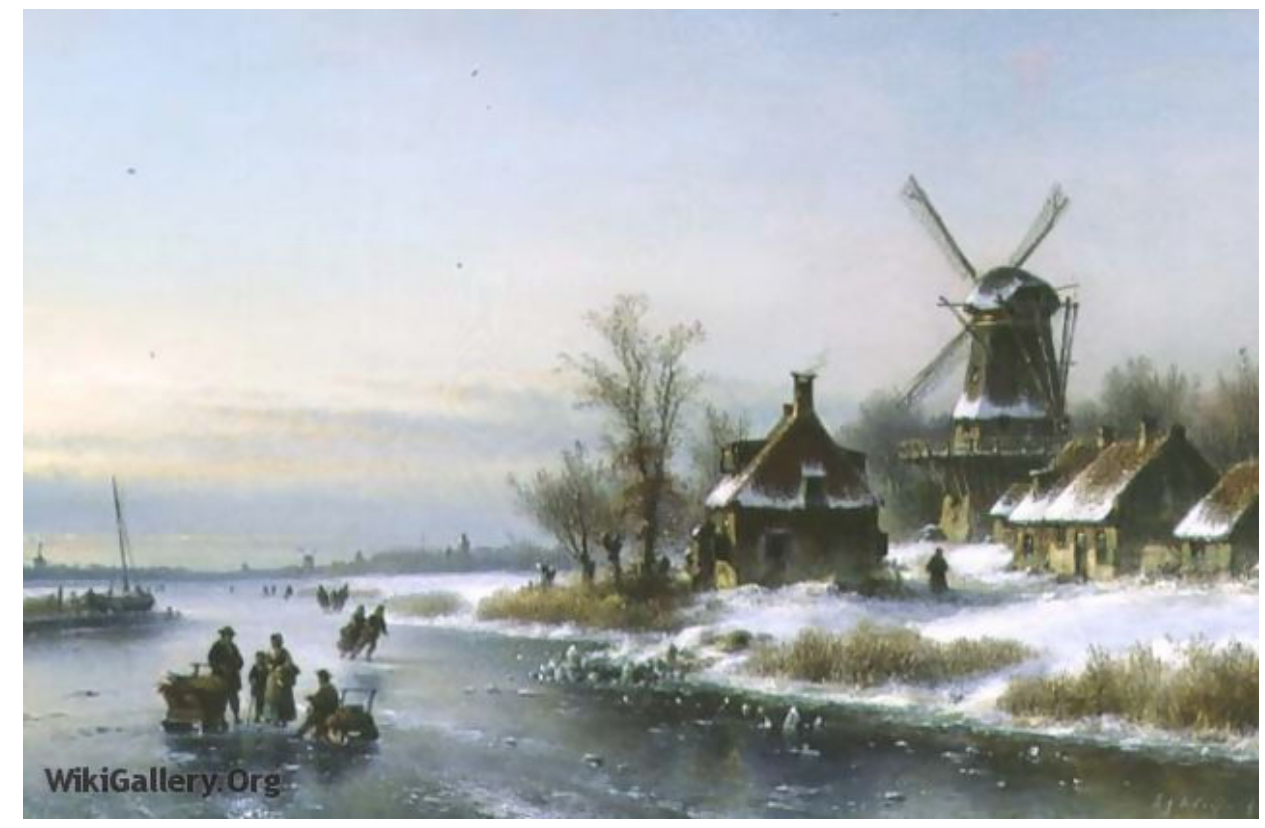

Fig. 5. "Windmill by a Frozen River" by Lodewijk Johannes Kleijn, evidencing weeks of freezing cold in Netherlands in the 19th century, already after the lowest point of the Dalton minimum, less intense than the Maunder minimum.

### C. *Climate change in this century*

Based on the evolution of the duration of the Schwabe cycle and on the satellite measurements of the total solar irradiance during the last 30 years, astronomers predicted the amplitudes of the next three Schwabe cycles, of about 70, 50 and 35 sunspots at the maximum, as compared to 110 of the last complete cycle. This means that a new Little Ice Age will be developing during the XXIst century, with the lowest solar activity achieved around 2040 and the temperatures hitting the bottom at around 2055, at some 1.0 K below the present value of the global average temperature, see Fig. 4 [21], [22]. This is comparable to what occurred during the Maunder minimum, being due to the existence of a bicentennial solar activity component, which produces changes in TSI that are significantly larger than those occurring during the Schwabe cycle, accompanied by increased intensity of GCR and increased cloud cover. This bicentennial component has produced notable climate differences, such as between the present solar maximum and the Maunder and Dalton minimums (Fig. 5), causing Little Ice Ages once in about every 200 years, as the historical record for the last 7500 years shows, with 18 of such cooling events documented [23].

## III. Social and Economic Consequences of Climate – Related Policies

AGW-motivated policies currently pursued in the EU on climate-related issues, including promotion of "green" fuels produced from plant matter, and large-scale installation of "green" energy production facilities, mostly for wind and solar energy harvesting, are already yielding disastrous consequences on a global scale, with zero possible impact on climate. Indeed, even the calculations made with IPCC's own data that exaggerate the [$CO_2$] effects upon the climate by a factor of 10, produce temperature reductions that could result from Kyoto and Copenhagen protocols amounting to several hundredths of a degree over this century, negligible compared to the astronomical costs of their implementation.

Promotion of the green fuels by developed countries has already caused a 75% increase of food prices on the world market, as concluded by a World Bank expert [24]. Given that between 10 and 20 million people die each year of famine-related causes [25], such a price hike has already caused and will continue causing millions of avoidable deaths. Adverse effects of atmospheric $CO_2$ inexistent, fuel production from plant matter is completely inexcusable: indeed, as we demonstrated recently, potential deposits of fossil fuels will be sufficient for at least one thousand years to come, and possibly many more [26].

Our estimate is based on the fact that oxygen ($O_2$) existing in the terrestrial atmosphere has been produced by plants, with the carbon-containing organic matter eventually buried underground and forming fossil fuels. The respective simplified chemical equation of the photosynthesis is:

$$\mathrm{CO_2}(g) + h\nu \rightarrow \mathrm{C}(s) + \mathrm{O_2}(g) \tag{1}$$

In Eq. 1 $h\nu$ represents the sunlight energy captured by plants, C the carbon they incorporated into organic matter, and $O_2$ the oxygen liberated into the atmosphere.

The amount of gaseous atmospheric $O_2$ corresponds to about 2000 kg/m$^2$ of the surface area of the globe, yielding about 800 kg/m$^2$ of equivalent carbon in fossil fuels. On the other hand, the quantity of fossil fuels existing in known and confirmed deposits of oil, coal, bituminous shale and natural gas corresponds to some 3 kg/m$^2$ [10]. This shows that potentially we have deposits of fossil fuels sufficient for 20000 years to come, which, after generous corrections made for difficulties in extracting some of these, allows us to conclude that we shall not be lacking fossil fuels for 1000 years, at the very least.

Another economically viable source of energy that can provide for all of the needs of the human civilization for the centuries to come are large liquid-metal cooled nuclear reactors on fast neutrons. These reactors are more robust than the more conventional designs and do not need enriched uranium or any other enriched radioactive materials, avoiding potential problems with proliferation of nuclear armaments [27]. They can be operated on the natural mixture of uranium isotopes or any other mixture of fissionable heavy isotopes, being very tolerant to the composition of the nuclear fuel. With proper design, these reactors may be build with a large safety margin, permitting the core to be cooled by gravitational convection of the coolant even in the absence of active pumping. Their additional advantage is the drastically reduced amount of nuclear waste as compared to the existing designs – potentially they can extract up to 98% of nuclear energy present in natural uranium, instead of the 2% of the existing designs. This also means that these reactors may be used to recycle almost all of the nuclear waste, already accumulated and to be produced by reactors of traditional design – indeed, after separation of the short-lived low atomic mass radioactive products, achieved by relatively simple and cheap chemical technologies, the remaining heavy radioactive isotopes may be recycled as nuclear fuel for such reactors, repeatedly and until consumed completely. With the drastically reduced total amount of nuclear waste, and with the long-living heavy isotopes all consumed as nuclear fuel, the problem of nuclear waste becomes much more manageable - apart from drastically reduced quantities, the required storage times until the waste reaches safe state is also drastically reduced.

On the other hand, the AGW-motivated drive for renewable energies is already producing very serious economic problems in Europe, including the current budget deficit in many of the EU countries, Portugal included. We have recently made estimates showing that half of the Portuguese state budget deficit may be explained by installation and usage of windmills for producing electric energy [28]. Indeed, the total installed windmill production capacity is about 3500 MW, with has capital costs of about 3900 million euros. The projected electric energy production on these installations for 2010 is about 9500 GWh, causing direct losses to the economy of about 470 million euros due to price differential of 50 €/MWh between conventional and windmill energy. The indirect economic losses due to increased production costs are much larger, totaling at least 2500 million euros. The actual losses are probably a double of this number, as we should add revenues lost due to reduced sales. With the about 40% overall taxation rates, the Portuguese state is therefore loosing at least 1200 million euros yearly in taxes, which together with tax breaks for green fuels, costs of carbon credits, incentives for green energy production and interest payments on loans contracted for renewable energies makes a very significant part of the Portuguese state budget deficit of 3700 million euros for the current fiscal year. Note that measures proposed to fight the budget deficit include salary cuts, tax hikes and cuts on public spending, directly affecting every taxpayer.

## IV. CONCLUSIONS

We demonstrated that the Anthropogenic Global Warming hypothesis and the IPCC climate projections/scenarios have nothing in common with the terrestrial climate system and thus must be expressly ignored when considering climate change and its consequences.

Anthropogenic carbon dioxide has contributed ca. 0.1 K to the climate change that occurred during the XXth century, and will contribute another 0.2 K in this century, provided its emissions continue unabated.

A simple phenomenological model, proposed by us and based on climate change patterns of the last 150 years, predicts global cooling to continue in the next 20 years.

Sun is the principal climate driver. The climate predicted for the XXIst century, based on the solar activity forecasts, is equivalent to a new Little Ice Age, with the lowest temperatures achieved by the middle of this century. Therefore, society should prepare a response to deal with global cooling that will be occurring in the next decades. In this respect, an urgent review of building codes is needed. This will prepare a significant percentage of homes to colder winters, when these come. In this context, any warming contribution of anthropogenic carbon dioxide will be a welcome factor stabilizing the climate, therefore any measures targeting reductions of $CO_2$ emissions should be immediately recalled, including green fuel production from plant matter and all of the economically unviable sources of renewable energies.

We should provide cheap and economically viable energy, including energy generated from fossil fuels, to all developing nations. Improved levels of life and well-being will reduce birth rates in the currently poor nations, alleviating pressures on the ecological systems. Increased atmospheric carbon dioxide concentrations will in turn significantly improve agricultural yields, by about 50% for doubled [$CO_2$], providing sufficient food at no additional cost for the growing world population [29].

## REFERENCES

[1] Intergovernmental Panel on Climate Change (2007), *Climate Change 2007: The Physical Science Basis*, report, 996 pp., Cambridge University Press, New York City, 2007.

[2] D. Xiao, Z. Zhang, H. Wang et al, "System Modeling for the Impact of Global Warming on Equity Price", 12th WSEAS International Conference on Computers, Heraklion, Greece, July 23-25, pp. 101-106, 2008.

[3] J.-M. Timmermans, J. Matheys, J. Van Mierlo and Ph. Lataire, "Ecoscore, an Environmental Rating Tool for Road Vehicles", in Proceedings of the 5th WSEAS International Conference on Environment, Ecosystems and Development, Venice, Italy, November 20-22, pp. 82-88, 2006.

[4] M. Tomosada, K. Kanefuji, Y. Matsumoto, and H. Tsubaki, "Application of the Spatial Statistics to the Retrieved $CO_2$ Column Abundances Derived from GOSAT Data", in Proceedings of the 4th WSEAS International Conference on Remote Sensing, Venice, Italy, pp. 67-73, 2008.

[5] R. Snow and M. Snow, "Climate Change Curricula and the Challenge for Educators", *WSEAS Trans. Envir. Devel.*, vol. 6, pp 54-58, 2010.

[6] R. Lindzen and Y.-S. Choi, "On the determination of climate feedbacks from ERBE data", *Geophys. Res. Lett.* vol. 36, pp. L16705, 2009. doi:10.1029/2009GL039628.

[7] R. W. Spencer, "Satellite and Climate Model Evidence Against Substantial Manmade Climate Change", available: http://www.drroyspencer.com/research-articles/satellite-and-climate-model-evidence/, *Journal of Climate*, submitted for publication.

[8] P. Corbyn, unpublished, available http://www.weatheraction.com/.

[9] N. A. Rayner, D. E. Parker, E. B. Horton, C. K. Folland, L. V. Alexander, D. P. Rowell, E. C. Kent, and A. Kaplan, Global analyses of sea surface temperature, sea ice, and night marine air temperature since the late nineteenth century, *J. Geophys. Res.* vol. 108, No. D14, p. 4407, 2003. doi:10.1029/2002JD002670. Hadley Centre SST data set HadISST1, available: http://hadobs.metoffice.com/hadisst/data/download.html

[10] Wikipedia, "Fossil Fuel", http://en.wikipedia.org/wiki/Fossil_fuel.

[11] SIDC-team, World Data Center for the Sunspot Index, Royal Observatory of Belgium, Monthly Report on the International Sunspot Number, online catalogue of the sunspot index, available: http://www.sidc.be/sunspot-data/, yrs 1700-2009.

[12] J. Lean, J. Beer, and R. Bradley, "Reconstruction of solar irradiance since 1610: implications for climate change", *Geophysical Research Letters,* vol. 23, pp. 3195–3198, 1995.

[13] J. Lean, Solar Irradiance Reconstruction. IGBP PAGES/World Data Center for Paleoclimatology, Data Contribution Series # 2004-035. NOAA/NGDC Paleoclimatology Program, Boulder CO, USA, 2004. Available: ftp://ftp.ncdc.noaa.gov/pub/data/paleo/climate_forcing/ solar_variability/ lean2000_irradiance.txt.

[14] E. Friis-Christensen and K. Lassen, "Length of the solar cycle: an indicator of solar activity closely associated with climate", *Science*, vol. 254, pp. 698-700, 1991.

[15] D. V. Hoyt and K. H. Schatten, "A discussion of plausible solar irradiance variations", *J. Geophys. Res.*, vol. 98, pp. 18895-18906, 1993.

[16] S. K. Solanki and M. Fligge, "Solar Irradiance Since 1874 Revisited", *Geophysical Research Letters*, vol. 25, no. 3, pp. 341-344, 1998.

[17] H. Svensmark, "Influence of Cosmic Rays on Earth's Climate", *Phys. Rev. Lett.*, vol. 81, pp. 5027–5030, 1998.

[18] H. Svensmark, J. O. P. Pedersen, N. D. Marsh, M. B. Enghoff and U. I. Uggerhøj, "Experimental evidence for the role of ions in particle nucleation under atmospheric conditions", *Proc. R. Soc.*, vol. A 463, pp. 385–396, 2007.

[19] H. Svensmark, "Cosmic Rays And Earth's Climate", *Space Science Reviews*, vol. 93, pp. 155–166, 2000.

[20] R. T. Pinker, B. Zhang and E. G. Dutton, "Do Satellites Detect Trends in Surface Solar Radiation?", *Science,* vol. 308. no. 5723, pp. 850-854, 2005. doi: 10.1126/science.1103159.

[21] H. Abdussamatov, "The Sun defines the Climate", 2008, not published, available: http://www.gao.spb.ru/english/astrometr/abduss_nkj_2009.pdf, http://www.gao.spb.ru/ english/astrometr/index1_eng.html.

[22] Kh. I. Abdusamatov, "Optimal Prediction of the Peak of the Next 11-Year Activity Cycle and of the Peaks of Several Succeeding Cycles on the Basis of Long-Term Variations in the Solar Radius or Solar Constant", *Kinematics and Physics of Celestial Bodies*, Vol. 23, No. 3, pp. 97–100, 2007.

[23] E. P. Borisenko, V. M. Pasetskii, *Climate variations during the last millennium*, Moscow, Mysl, 1988. Available: http://www.pereplet.ru/gorm/dating/ climat.htm

[24] D. Mitchell, "A Note on Rising Food Prices", online, available: http://www-

wds.worldbank.org/external/default/WDSContentServer/IW3P/IB/2008/07/28/000020439_20080728103002/Rendered/PDF/WP4682.pdf
[25] G. Polya, "The Unnecessary Bengali Famine", *The Muslim Observer*, 2007, online, available: http://muslimmedianetwork.com/mmn/?p=868.
[26] I. Khmelinskii, "Fossil fuels are dwindling - you got to be kidding!", unpublished, online, available: http://clima-virtual-vs-real.blogspot.com/2010/10/fossil-fuels-are-dwindling-you-got-to.html.
[27] Wikipedia, "Liquid metal cooled reactor", online, available: http://en.wikipedia.org/wiki/Liquid_metal_cooled_reactor.
[28] I. Khmelinskii, "As Consequências Económicas de Energias Renováveis: Caso de Portugal", unpublished, online, available: http://clima-virtual-vs-real.blogspot.com/2010/10/as-consequencias-economicas-de-energias.html.
[29] T.J Blom, W.A. Straver, F.J. Ingratta, S. Khosla and W. Brown "Carbon Dioxide In Greenhouses", Canadian Ministry of Agriculture, Food and Rural Affairs, online, available: http://www.omafra.gov.on.ca/english/crops/facts/00-077.htm.

**Igor Khmelinskii** was born in Sverdlovsk, USSR, in 1957. Got his degree in Physics, majoring in Chemical Physics, from the Novosibirsk State University in 1979, PhD in Chemical Physics from the Institute of Chemical Kinetics and Combustion, Siberian Branch of the USSR Academy of Sciences, in 1988, and Habilitation in Physical Chemistry from the Universidade do Algarve, in 2003.

He is affiliated to Universidade do Algarve, Faro, Portugal from 1993, teaching Physical and General Chemistry at all levels, having previously worked as a Research Assistant in Novosibirsk, Russia. His research interests include Photochemistry, Spectroscopy and Applications, Magnetic Field Effects and Climate Change.

Mr. Khmelinskii is not a member of any professional society.